\documentstyle [ emulateapj]{article}

\newcommand{\lya}{Ly$\alpha$ }

\newcommand{\dd}{\,{\rm d}}

\newcommand{\hii}{\mbox{H\,{\scriptsize II}\ }}

\newcommand{\heii}{\mbox{He\,{\scriptsize II}\ }}
\newcommand{\heiii}{\mbox{He\,{\scriptsize III}\ }}
\newcommand{\op}{Ly$\alpha$\ }
\def\expec#1{\langle#1\rangle}

 1

\def\kms{\,{\rm {km\, s^{-1}}}}

\def\etal{{et al.~}}

\def\erg{{\rm erg}}
\def\sr{{\rm sr}}

\def\gs{\mathrel{\raise1.16pt\hbox{$>$}\kern-7.0pt
\lower3.06pt\hbox{{$\scriptstyle \sim$}}}}
\def\ls{\mathrel{\raise1.16pt\hbox{$<$}\kern-7.0pt
\lower3.06pt\hbox{{$\scriptstyle \sim$}}}}
\def\gtsima{$\; \buildrel > \over \sim \;$}
\def\ltsima{$\; \buildrel < \over \sim \;$}
\def\prosima{$\; \buildrel \propto \over \sim \;$}
\def\gsim{\lower.7ex\hbox{\gtsima}}
\def\lsim{\lower.7ex\hbox{\ltsima}}
\def\simgt{\lower.7ex\hbox{\gtsima}}
\def\simlt{\lower.7ex\hbox{\ltsima}}
\def\simpr{\lower.7ex\hbox{\prosima}}
\def\la{\lsim}

\def\pp{\noindent\parshape 2 0truecm 17truecm 2truecm 15truecm}
\def\rf#1;#2;#3;#4 {\par\pp#1, #2, #3, #4. \par}

\def\pr{\ref@jnl{Phys.Rev}}

\def\href#1;#2 {{\bf #1} : {\em #2}}

\def\beq#1{\begin{equation}\label{#1}}
\def\eeq{\end{equation}}
\def\beqa#1{\begin{eqnarray}\label{#1}}
\def\eeqa{\end{eqnarray}}
\def\eq#1{equation~(\ref{#1})}

\def\tento#1{\times 10^{#1}}

\def\K{{\rm \ K}}
\def\s{{\rm \ s}}
\def\sr{{\rm \ sr}}

\def\erg{{\rm \ erg}}

\def\Mpc{{\rm \ Mpc}}

\def\Hz{{\rm \ Hz}}

\def\H2p{H$_2^+$ }

\def\mH2p{H_2^+}

\makeatletter

\newenvironment{figurehere}
  {\def\@captype{figure}}
  {}
\makeatother

\begin{document}
\textheight=24.5cm

\title {Radiative transfer effects during photoheating 
of the Intergalactic Medium}
\author {Tom Abel\altaffilmark{1,2} and  Martin G. Haehnelt\altaffilmark{2}}
\received{soon}
\accepted{later}

\altaffiltext{1}{ Laboratory for Computational Astrophysics, NCSA,
          University of Illinois at Urbana/Champaign, 405 N. Mathews
          Ave., Urbana, IL 61801.}  \altaffiltext{2}{
          Max-Planck-Institut f\"ur Astrophysik,
          Karl-Schwarzschild-Strasse 1, 85748 Garching, Germany }

\begin{abstract}
  The thermal history of the intergalactic medium (IGM) after
  reionization is to a large extent determined by photoheating.  Here
  we demonstrate that calculations of the photoheating rate which
  neglect radiative transfer effects substantially underestimate the
  energy input during and after reionization.  The neglect of
  radiative transfer effects results in temperatures of the IGM which
  are too low by a factor of two after \heii reionization. 
  We briefly discuss implications for the absorption properties of 
  the IGM and  the distribution of baryons in shallow potential 
  wells.\\ 
\end{abstract}
\section{Introduction}
\thispagestyle{empty}

The absence of a Gunn-Peterson trough (Gunn \& Peterson 1965, Scheuer 1965) 
in the spectra of high-redshift objects is solid evidence that the Universe 
has been reionized before $z\sim5$.
Hydrodynamical simulations have shown convincingly that at redshifts
$z\sim 3$  most baryons are still contained in a photoionized IGM 
which is responsible for the \lya
forest in the absorption spectra of high-redshift QSO's (Cen et
al.~1994, Petitjean, M\"ucket \& Kates 1995; Zhang, Anninos \& Norman
1995; Hernquist et al.~1996).  However, the detailed distribution of the 
baryonic component as well as the absorption properties of the IGM will 
depend on the thermal history of the IGM and thus the heat input due to 
photoheating (Efstathiou 1992, Miralda-Escude \& Rees 1994, Navarro \& 
Steinmetz 1997, Haehnelt \& Steinmetz 1998, Bryan~\etal~1999; 
Theuns~\etal~1999). Most numerical simulations so far adopted the optically 
thin limit when calculating
photoheating rates.  This is, however, a bad approximation
during  reionization when the optical depth is large. 
As we will argue later in the case of
helium it is also a bad approximation after \heii
reionization (see also Giroux \& Shapiro 1996). 
Here we investigate radiative transfer  effects during and 
after reionization and discuss implications for the thermal 
history and absorption properties of the IGM and the distribution of 
baryons in shallow potential wells. We assume an
Einstein-de-Sitter Universe and a Hubble constant 
of $50\kms \Mpc^{-1}$.

\section{Photoheating}

\subsection{Optical thin {\it vs} optical thick}

In the optically thin limit the mean excess energy 
of ionizing photons is given by 
\begin{eqnarray}
  \label{eq:av_eg}
    \langle E_{\rm ph}^{\rm thin} \rangle = {{\int_{\nu_{\rm th}}^\infty
  \frac{J(\nu)}{h\nu}\sigma (\nu) (h\nu-h\nu_{th}) d\nu}\over
  {\int_{\nu_{\rm th}}^\infty
  \frac{J(\nu)}{h\nu} d\nu}}
\end{eqnarray}
where $J(\nu)$ is the flux of ionizing photons $\sigma(\nu)$ is the
crossection for ionization, $\nu_{th}$ the threshold energy for
ionization, and $h$ Planck's constant.  For a powerlaw spectrum
$J(\nu) = J_0\,(\nu/\nu_{th})^{-\alpha}$ and a powerlaw approximation
($\propto \nu^{-3}$) for the crossection one finds $\langle E_{\rm
  ph}^{\rm thin} \rangle \approx h\nu_{\rm th}/(\alpha+2)\equiv E_{\rm
  th}/(\alpha+2)$. For spectral indices of interest $\alpha >1$ these
excess energies are small fractions of the threshold energy $E_{\rm
  th}$.  This is because the integral in \eq{eq:av_eg} is strongly
weighted towards the threshold energy. In the optical thick limit
where {\it every}\, ionizing photon emitted is absorbed the mean
excess energy is typically much larger than $\langle E_{\rm ph}^{\rm
  thin} \rangle$ of \eq{eq:av_eg}. For the power-law spectrum
discussed above $\langle E_{\rm ph}^{\rm thick} \rangle \approx
h\nu_{\rm th}/(\alpha-1)$.

\subsection{Energy input during reionization} 

A non--equilibrium multi--frequency 1-D radiative transfer code is used to
study the thermal structure of R-type ionization fronts surrounding a
point source in spherical symmetry (Abel, Meiksin, \& Norman 1999).
The code evolves the cooling/heating and chemistry model for primordial
gas of Abel \etal (1997) with the methods discussed by Anninos \etal
(1997). It solves the static radiative transfer equation in an
expanding universe and explicitly conserves photons (Abel,
Norman and Madau 1999). 

Figure~1 shows radial temperature profiles of two different ionization
fronts around a luminous QSO (photon emission rate
$N_{\rm ph}=3\tento{56}\s^{-1}$, $\epsilon (\nu) =
\epsilon_0\,(\nu/\nu_0)^{-1.8}$) at $z=6$ and $z=4$, respectively.  In
both cases the front evolves into a homogeneous IGM of primordial
composition and mean baryonic density ($\Omega_{\rm bar} h^2= 0.02$).
In the first case the IGM is assumed to be completely neutral
and hydrogen and helium reionization occur almost simultaneously. In
the second case the gas is assumed to be predominantly in the form of
HII and  HeII due to a soft UV background already
present when the QSO switches on.  For both models the ionization
front is shown after $10^{7}$ and $10^{8}$ yr.

In the first model the ionization front is very sharp due 
to the small mean free path
of hydrogen ionizing photons and the temperature behind the
reionization front rises  to 50000 K. This corresponds to 
$\expec{E_{\rm ph}} \sim 0.85 \times E_{\rm th}$. 
This is much larger than in the optical thin approximation, 
$\expec{E_{\rm ph}^{\rm thin}}= 0.26 E_{\rm th}$,
and approaches the optical thick limit, 
$\expec{E_{\rm ph}^{\rm thick}}= 1.25 E_{\rm th}$, for the adopted 
power-law spectrum with $\alpha = 1.8$.  The difference is mainly due 
to line-cooling of the neutral component in the ionization front.
This becomes more apparent at later times when the ionization front
has slowed down.  Hydrogen line cooling of the neutral
component during the passage of the ionization front becomes 
then more effective and the temperature falls to 28000 K 
(Miralda-Escude \& Rees 1994). This is still  nearly a 
factor two larger than  the value of $15000 K$ in the optical 
thin limit (e.g. Haehnelt \& Steinmetz 1998). 

\vspace{+0.2cm}
\begin{figurehere}
\plotone{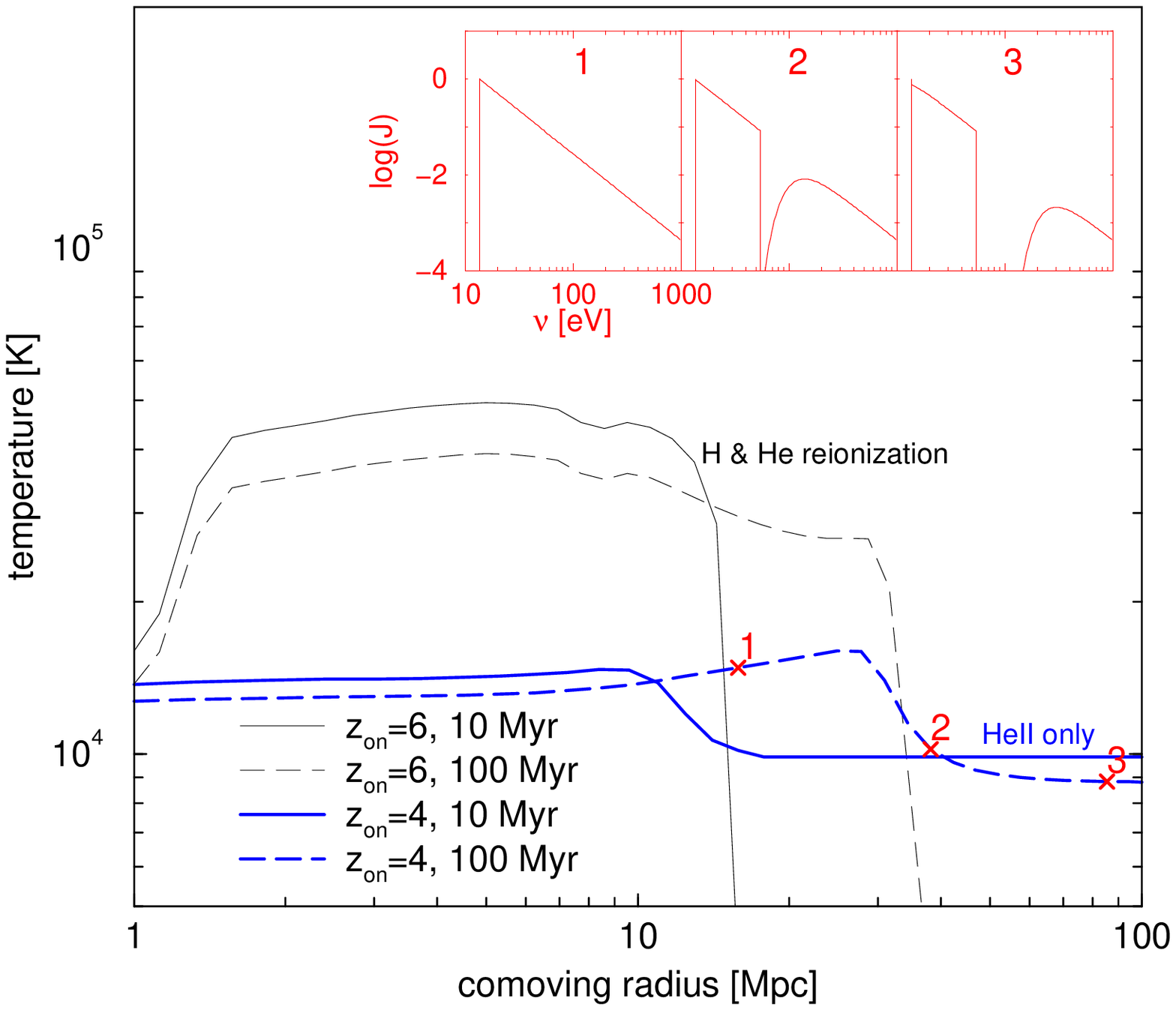}\vspace{-0.4cm}
\caption{ \footnotesize 
Thin and thick curves show the radial temperature profiles 
of the ionized region around a luminous quasar
($N_{\rm ph}=3\tento{56}\s^{-1}$)  that switched on at $z=6$ and 
$z=4$, respectively. In the first case the ionization front evolves 
into a completely neutral homogeneous IGM  of primordial composition
while in the latter case the gas  is assumed to be predominatly  
in the form of \hii and \heii.
For the second case the inset shows the spectral shape 
of the ionizing flux at the three different radii marked in the
figure.}\label{T(r)}
\end{figurehere}
\vspace{0.1cm}

In the second case  only helium is reionized and  hydrogen line cooling   
plays no role due to the small neutral fraction of hydrogen. 
The mean free path of HeII ionizing photons is about a factor
fifty larger  than that of hydrogen ionizing photons 
and the ionization front extends now over a few (comoving) Mpc. 
The temperature is raised from 9000K to about 17000K 
and the total energy input corresponds 
to $\expec{E_{\rm ph}} \sim  0.5 \times E_{\rm th}$, but 
this value rises to $\expec{E_{\rm ph}} \sim 0.7 \times E_{\rm th}$
when the energy deposited ahead of the front is taken into account. 
The  value is again smaller than that in the optical thick 
limit. This time because the mean free path of an helium ionizing 
photon becomes larger than the size of the ionized  region for 
$(1+E_{\rm ph}/E_{\rm th})> 3.8\,(R/30\Mpc)^{1/3}$,
where we have assumed that the ionization cross section scales as
$\nu^{-3}$ above the threshold. This also explains why the IGM 
temperature behind the ionization front increases with increasing
radius. The inset in Figure  1 shows the spectral shape behind, within
and ahead of the front to demonstrate how the spectrum  of absorbed 
photons becomes harder across the ionization front.

\subsection{Photoheating after reonization}

After reionization the mean free path of hydrogen ionizing photons is
comparable to the mean distance between hydrogen Lyman limit systems
(LLS) which is about $350 ((1+z)/4)^{-3} \Mpc $ comoving
(Storrie-Lombardi 1994). Much less is known about the number of helium
LLS.  The model of Miralda-Escud\'e, Haehnelt, \& Rees (1999, MHR99)
which uses a realistic density distribution to treat inhomogeneous
reionization predicts a typical distance of $\sim 30 \Mpc$ 
around $z\sim 3$ (see also Fardal, Giroux  \& Shull 1998).
Both values are larger than the typical size
of numerical simulations and this is is often taken as argument to
treat photoheating in the optically thin limit. However, it is not
before z=1.6 that the mean free path of hydrogen ionizing photons
becomes comparable to the Hubble radius (e.g.  Madau, Haardt \& Rees
1999) and the Universe is likely to be opaque to helium ionizing
photons up to the present day.  Almost all photons emitted at high
redshift will therefore be absorbed within a Hubble time and
calculations adopting the optically thin limit must severely
underestimate at least the mass-weighted average photoheating rate
after reionization.

In the case of hydrogen the mean free path of an hydrogen ionizing
photon is smaller than the distance between ionizing sources and the
optical depth of LLS is large. At a typical place in the Universe the
UV background will be dominated by ionizing photons emitted within the
region filling the volume to the next LLSs.  In the case of HeII the
mean distance between sources of ionizing photons at $z\sim 3$ is
$\sim 100 \Mpc $ (MHR99) and thus still larger than the mean distance
between helium LLS.  At a typical point in the Universe the HeII
ionizing photons will thereforehave passed a helium LLS. The minimum
energy input per ionization is thus raised such that $(1+E_{\rm
  ph}/E_{\rm th})> \tau ^{1/3}$, where $\tau$ is the optical depth at
the ionization threshold.  Little is known about the optical depth
distribution of Helium LLS but even a rather low optical depth of
$\tau \sim 10$ would raise the HeII photoheating rate by more than a
factor of four compared to the optical thin case.

However, even when the distance between helium LLS falls below the
mean distance between sources of ionizing photons the optical thin
approximation will significantly underestimate the HeII photo-heating
rate, unless helium LLS have a very large optical depth. To understand
this consider the extreme case of a homogeneous emissivity of ionizing
photons and many helium LLS with optical depth of order unity at the
\heii ionization threshold.  The ionizing background at a given
frequency would then be given by the emisssivity times the mean free
path at this frequency, $I(\nu) \propto \epsilon(\nu) l(\nu) \propto
\epsilon(\nu) \sigma(\nu)^{-1}$.  In this limit the mean absorbed
photon energy is equal to the mean emitted energy of photons.

An accurate prediction of the mean absorbed energy is extremely
difficult as it depends very sensitively on the spatial distribution
of HeII optical depths and ionizing sources during and after
reionization and on their emitted spectrum. We can here only roughly
estimate that $E_{\rm ph}/ E_{\rm th} \sim$ 0.5 to 1 at $z\sim 3$ and
should decrease with redshift. The effective HeII photoheating rate
should thus be between 2 and 4 times larger than that in the optical
thin approximation. It could be even larger, if the emitted spectrum
were harder than the $\alpha =1.8$ powerlaw assumed here.

\vspace{0.2cm}
\begin{figurehere}
\plotone{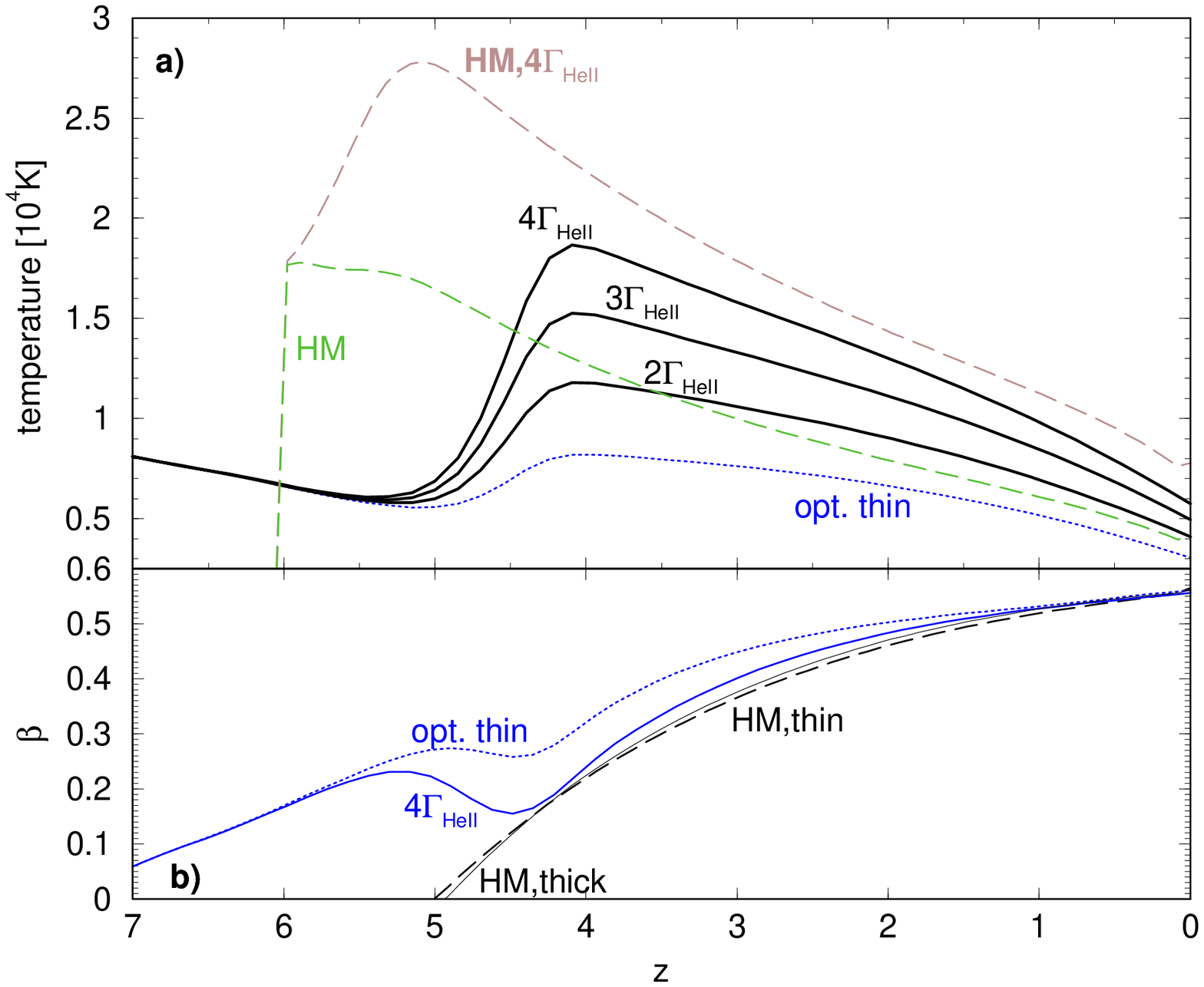}\vspace{-0.4cm}
\caption{ \footnotesize 
  a) The temperature evolution of an IGM of primordial composition and
  mean density assuming a homogeneous UV background as in equation
  (3).  For the solid curves the HeII photoheating rate was raised by
  a factor 2,3,4, respectively. The dashed curves assume 
  a homogeneous UV background as given by Haardt \& Madau (1996)
  with the HeII photoheating for  the optical thin limit 
  and raised by a factor four, respectively.  
   b) The slope $\beta = \dd \ln{T}/\dd \ln{\rho}$ of the temperature
   density relation  at mean density.}   
\end{figurehere}

\section{Implications}

\subsection{Temperature of the IGM}

In the last section we have demonstrated that optical depth effects 
raise the effective helium photoheating rate by about a factor 2 to 4. 
Here we investigate the effect of such an increased helium
photoheating rate on the temperature of the IGM. We used a 
non-equilibrium code to calculate the redshift evolution of the
ionization state and temperature of an photoionized IGM of primordial 
composition with mean baryonic density. We have assumed 
a homogeneous  UV background of the form, 
\begin{eqnarray}
I(\nu,z) &=&  
       \left[ \frac{1}{1 +(7/(1+z))^4} e^{-(z/4)^3}
       \left(  \frac{\nu}{\nu_{\rm HI}} \right)^{-5} \right. \nonumber\\  
         &+&    \left.
 \frac{10}{1+(7/(1+z))^4} e^{-(z/2.5)^3}
\left( \frac{\nu}{\nu_{\rm HI}}\right)^{-1.8} \right] \nonumber\\
&\times & 10^{-21} \erg \s^{-1} \Hz^{-1} \sr^{-1}. 
\end{eqnarray} 
This is similar to model C in Haehnelt \& Steinmetz 1998.  The first
term mimicks reionization of hydrogen by stars before $z=6$. The
second component represents the contribution of QSO's to the UV
background which here results in a separate epoch of HeII
reionization around $z=4$.

The dotted curves shows the temperature evolution for the optical thin
case. For the solid curves we have raised the HeII photoheating rate
by factors 2,3 and 4.  This corresponds to an increase in the {\it total}
photoheating rate by a factor 1.3, 2 and 2.6, respectively.  For
reference the dashed curves show the temperature evolution for the UV
background as calculated by Haardt \& Madau (1996) in the optical thin
limit and with the HeII photoheating rate increased by a factor of four.

As pointed out by by MHR99 the assumption of an homogeneous UV
background during reionization results in an artificially abrupt
reionization. In reality reionization should occur over a significant
fraction of the Hubble time.  The assumption of a homogeneous
background during reionization should nevertheless give a resonable
estimate of the mean temperature after reionization even though there
will be considerable spatial temperature fluctuations.
 
At $z\la 6$ adiabatic cooling exceeds Compton cooling and after helium
reionization and in a fully photoionized gas the temperature is thus
determined by the balance between the photoheating rate, $\Gamma_{\rm
  ph} = (\alpha_{\rm HeIII} \expec {E_{\rm ph}^{\rm HeIII} }n_{\rm
  HeIII} + \alpha_{\rm HII} \expec {E_{\rm ph}^{\rm HII}} n_{\rm HII})
n_{\rm e}$ and the adiabatic cooling rate $C_{\rm ad} = 3 kT H(z) \rho
/ \mu m_{\rm p} $, where $\alpha (\propto T^{-0.7})$ is the
recombination coefficient , $n_e$ is the electron density, $H(z)$ is
the Hubble constant $\mu$ is the mean molecular weight and $m_{\rm p}$
is the proton mass.  Note that there is no explicit dependence on the
strength of the radiation field.  This results in a quasi-equilibrium
temperature which is approached at late times (Miralda Escud\'e \&
Rees 1994)
\begin{eqnarray}\label{eos}
T_{\rm ad} &\sim&  5.9\tento{3} \;
\left( {1+z} \right)^{\frac{1.5}{1.7}} 
\left( \frac{\rho} {\expec{\rho}} \right)^{\frac{1}{1.7}} \nonumber
\left( \frac{\Omega_{\rm b} h}{0.04} \right)^{\frac{1}{1.7}}  \\
& & 
\left(\frac{\expec{E_{\rm ph}^{\rm HeIII}}}{E_{\rm th}^{\rm HeIII}}
+ 0.15  \right )^{\frac{1}{1.7}}        
{\rm K },
\end{eqnarray}
where  we have taken the optical thin limit 
for hydrogen heating. The increased HeII photoheating rate 
due to optical depth effects
results in an increase of the temperature by factors 1.5--2.5
(Fig. 2a). Changing to a cosmological model with $\Omega =0.3$ would further
raise this temperature by 20 percent at redshift three because of the 
decrease in the adiabatic cooling rate. 

Madau \& Efsthathiou (1999) recently suggested that Compton
heating by the hard X-ray background might significantly contribute 
to the heating of the IGM. They  obtain 
for the  Compton heating rate at the present day 
$H_{\rm co} = 1.25\times 10^{-31} \erg \s^{-1}$.
We can now compare that to the  photoheating rate 
\begin{eqnarray}\label{ratio} 
\frac{\Gamma_{\rm ph}}{H_{\rm co} f(z) n_{e}}  & \sim&  23 \,
\left(\frac{T}{10^4  \K }\right )^{-0.7}  
\left(\frac{\Omega_{\rm bar}h^2}{0.02}\right ) \nonumber \\ 
&\times&
\left (\frac{\expec{E_{\rm ph}^{\rm HeIII}}}{E_{\rm
th}^{\rm HeIII}} + 0.15 \right )   
\frac{\rho}{\expec{\rho}}  
\frac{(1+z)^{3}}{f(z)}. 
\end{eqnarray}
Note that the photoheating rate incrases with density while the
Compton heating rate does not.  The redshift evolution of the Compton
heating rate $f(z)$ depends on the evolution of the energy density of
the hard X-ray background the mean energy of the Compton scattered
photon.  The Klein-Nishina reduction to the Compton cross section
becomes increasingly important at high redshift. Madau \& Eftathiou
(1999) obtain $f(z) \sim (1+z)^{13/3}$ as approximate redshift
dependence if the hard X-ray background was produced at high redshift.
Hence, the Compton heating rate by the hard X-ray background exceeds
the photo-heating rate only at redshifts larger than
\begin{eqnarray}\label{red} 
(1+z)& \sim&  10 \,
\left(\frac{T}{10^4  \K }\right )^{-0.52}  
\left(\frac{\Omega_{\rm bar}h^2}{0.02}\right )^{3/4} \nonumber \\ 
&\times&
\left (\frac{\expec{E_{\rm ph}^{\rm HeIII}}}{E_{\rm
th}^{\rm HeIII}} + 0.15 \right )^{3/4}   
\left (\frac{\rho}{\expec{\rho}} \right ) ^{3/4}.   
\end{eqnarray}

\subsection{Absorption properties of the IGM} 

The raise in temperature should affect the absorption properties of
the IGM in several ways. It will increase the thermal broadening of
absorption lines but might also introduce additional non-thermal
broadening due to pressure-induced motions.  The temperature increase
due to optical depth effects during HeII photoheating is therefore a
good candidate to resolve recently reported difficulties of numerical
simulations to reproduce Doppler parameters of \lya absorption lines
at $z=$ 2 to 3 as large as those observed (Bryan~\etal~1999;
Theuns~\etal~1999).  This would otherwise require an unreasonable
large value of $\Omega_{\rm b} h^2$ (Theuns et al 1999).  The
temperature increase will also increase the mean baryonic density
required to match the mean flux decrement in QSO absorption spectra
(Rauch et al. 1997).  The details of helium heating will furthermore
affect the temperature density relation (Hui \& Gnedin 1997) and thus
the flux decrement distribution.  To demonstrate the effect on the
temperature density relation we have plotted in Fig. 2b the
logarithmic slope $\beta = \dd \ln{T}/\dd \ln{\rho}$ at mean density
for the same models as in Fig 2a.

\subsection{Baryons in shallow potential wells}
The increased helium photoheating rate will also raise the Jeans mass 
and will decrease the ability of the gas to cool and collapse 
into shallow potential wells. It will thus affect the overall 
distribution of baryons. 
For  photoionized gas the dominant cooling processes during 
the early stages of the collapse are recombination 
cooling and Bremstrahlung.  The equilibrium between photoheating 
and  recombination cooling ($C_{\rm rec} \propto \rho^2  T^{0.3}$) 
results in a temperature 
$T_{\rm rec} \sim  2.3 \tento{5} \;
[ ({\expec{E_{\rm ph}^{\rm HeIII}}/E_{\rm th}^{\rm HeIII}})
+ 0.15  ]        
{\rm K }$. 
The increased HeII photoheating rate should thus affect the circular
velocity threshold above which gas can collapse into dark matter
haloes and the fraction of gas that can cool in shallow potential
wells (Thoul\& Weinberg 1996, Forcada-Miro 1998). Similarly,
the increased hydrogen and helium photoheating rates should raise the
circular velocity threshold below which baryons in shallow potential
wells which collapsed before reionization can be evaporated by an
ionization front (Shapiro \etal~1998; Abel \& Mo 1998; Barkana \& Loeb
1999).

\section{Conclusions}

We have studied radiative transfer effects on the energy input into
the IGM due to the reionization and photoheating of hydrogen and
helium. The energy input depends crucially on the spectral
distribution of the photons actually absorbed. 

During reionization the spectrum of an ionizing source becomes harder 
across the ionization front. The energy input  is close 
to the mean  energy of the photons between the ionization threshold and the
frequency for which the mean free path of an ionizing photons becomes
larger than the radius of the ionized region. 

After the reionization of HeII the mean spectrum of ionizing photons
above the \heiii ionization edge will also be considerably harder than
the emitted spectrum of an individual source.  This is because hard
photons can pass through helium LLS.  The number and optical depth
distribution of helium LLS is still highly uncertain. Thus their
effect on the UV specrtum is likely to have been underestimated
previously.

The increased HeII photoheating rate due to optical depth effects
raises the temperature of the IGM typically by about a factor 1.5 to
2.5 compared to the optical thin approximation.  It might thus resolve
the problem that current numerical simulations of the \op forest
(which treat photoheating in the optical thin limit) produce Doppler
parameters narrower than observed.

The increased HeII photoheating rate will reduce the ability of the
gas to collapse into shallow potential wells. It will also raise the
circular velocity threshold below which gas in dark matter haloes
is photoevaporated during reionization.

\acknowledgments Tom Abel is grateful to Avery Meiksin and Michael
Norman for allowing us to show first results from the jointly
developed 1--D radiative transfer code. We also thank Greg Bryan,
Martin Rees, Tom Theuns and Simon White for stimulating discussions
and suggestions. This work was partially supported by NASA through ATP
grant NAG5-4236. Tom Abel acknowledges support from NASA grant
NAG5-3923.


\vfill
\eject
\end{document}